\documentclass[preprint,3p,times]{elsarticle}
\usepackage{graphicx}
\usepackage{amssymb}
\usepackage{amsmath}
\usepackage{bm}
\usepackage{hyperref}

\usepackage{color}
\usepackage[T1]{fontenc} % if needed

\newcommand{\kperp}{\boldsymbol{k}_T}

\newcommand{\bT}{b_T}
\newcommand{\bb}{b}

\newcommand{\kperpn}[1]{}

\begin{document}

\title{The 3-dimensional distribution of quarks in momentum space}
\tnotetext[t1]{{\it Preprint number:} JLAB-THY-20-3186}

\author[1]{Alessandro Bacchetta}
\ead{alessandro.bacchetta@unipv.it}
\address[1]{Dipartimento di Fisica, Universit\`a di Pavia,
  and INFN Sezione di Pavia, via Bassi 6, I-27100 Pavia, Italy}

\author[2]{Filippo Delcarro}
\ead{delcarro@jlab.org}
\address[2]{Jefferson Lab, 12000
  Jefferson Avenue, Newport News, VA 23606, USA}

\author[3]{Cristian Pisano}
\ead{cristian.pisano@unica.it}
\address[3]{Dipartimento di Fisica, Universit\`a di Cagliari,
  and
  INFN Sezione di Cagliari, Cittadella Universitaria, I-09042
  Monserrato (CA), Italy}

\author[4]{Marco Radici}
\ead{marco.radici@pv.infn.it}
\address[4]{INFN Sezione di Pavia, via Bassi 6, I-27100 Pavia, Italy}

%\address[1]{Dipartimento di Fisica, Universit\`a di Pavia,
%  and INFN Sezione di Pavia, via Bassi 6, I-27100 Pavia, Italy}
%\address[2]{Jefferson Lab, 12000
%  Jefferson Avenue, Newport News, Virginia 23606, USA}
%\address[3]{Dipartimento di Fisica, Universit\`a di Cagliari,
%  and
%  INFN Sezione di Cagliari, Cittadella Universitaria, I-09042
%  Monserrato (CA), Italy}
%\address[4]{INFN Sezione di Pavia, via Bassi 6, I-27100 Pavia, Italy}

\begin{abstract}
We present the distribution of unpolarized quarks in a transversely
polarized proton in three-dimensional momentum space. Our results are
based on the extractions of the unpolarized and Sivers transverse
momentum dependent parton distributions (TMDs) in a fully consistent TMD framework.
%on consistent extractions of the unpolarized and Sivers transverse
%momentum dependent parton distributions (TMDs). 
\end{abstract}

%\date{\today, \currenttime}

%%%%%%%%%% MARCO
%%  mettere la data di prima sottomissione e poi quella revised?
%%%%%%%%%%

%\preprint{JLAB-THY-20-3186}

\maketitle

%\textbf
%\section{Introduction}
{The antipode of taking a picture of a black hole is to take a picture of the
inside of a proton, unveiling its internal constituents, confined in the
most common element of the visible universe by the strong forces of Quantum
Chromodynamics (QCD). 
Using data obtained from the scattering of a hard virtual photon off a proton, we map
the density of quarks in three dimensions, i.e., as a function of their
longitudinal momentum (along the photon's direction) and their transverse
momentum (orthogonal to the photon). If the proton is unpolarized, the
distribution is cylindrically symmetric: we determine it using recent results
from our group~\cite{Bacchetta:2017gcc}.  
If the proton is polarized in the transverse plane, the distributions of up
and down quarks turn out to be distorted in opposite directions. This
distortion, known as Sivers effect~\cite{Sivers:1989cc}, is related to quark
orbital angular momentum. We determine its details with the same formalism
used for the unpolarized distribution. In this way, we obtain a consistent
picture of the full 3-dimensional momentum
distribution of quarks in a transversely
polarized proton.
Our study constitutes a benchmark for future determinations of
multi-dimensional quark distributions, one of the main goals of existing and
planned experimental
facilities~\cite{Boer:2011fh,Accardi:2012qut,Dudek:2012vr,Aschenauer:2016our}.  
}

%\section{Formalism}
We consider a frame where the proton has momentum $P$ with space component in
the $+\hat{z}$ direction, is polarized in the $+\hat{y}$ direction, and is probed by
a spacelike
virtual photon with momentum $q$ (with $Q^2 = -q^2$) in the $-\hat{z}$ direction.
We define the $\widehat{xy}$ plane as transverse and we denote it with the subscript
$T$. We consider the light-cone $+$ direction $(\hat{t}+\hat{z})/\sqrt{2}$ and we
define it as longitudinal. If $Q^2$ is much larger than the
proton's mass $M^2$,
the proton's momentum is approximately longitudinal ($P^+$ 
is the dominant component). 

Our goal is to reconstruct the distribution of unpolarized quarks inside the
nucleon as a function of three components of their momentum. 
In the frame we are considering, 
the distribution of a quark with flavor $a$ in a transversely polarized
nucleon $N^\uparrow$ can be written in terms of two
Transverse Momentum Distributions (TMDs) as~\cite{Bacchetta:2004jz}
\begin{equation}
\begin{split}
\rho^a_{N^\uparrow}(x, k_x, k_y; Q^2)
 &= f_1^a (x, k_T^2;Q^2) -\frac{k_x}{M} 
f_{1T}^{\perp a}(x, k_T^2;Q^2)\, ,
\end{split}
  \label{e:density}
\end{equation}
where $f_1^a$ is the unpolarized TMD and $f_{1T}^{\perp a}$
is the Sivers TMD~\cite{Sivers:1989cc}, $k$ is the
momentum of the quark, $k_T$ the modulus of its transverse component, and $x=k^+/P^+$ is its
longitudinal momentum fraction. $Q^2$ plays the role of a resolution scale.

Recent extractions of $f_1$ have been published in
Refs.~\cite{Bacchetta:2017gcc,Bertone:2019nxa,Scimemi:2019cmh,Bacchetta:2019sam}. Several
parametrizations of $f_{1T}^\perp$ have been released up to now~\cite{Anselmino:2005ea,Collins:2005ie,Vogelsang:2005cs,Anselmino:2008sga,Bacchetta:2011gx,Sun:2013hua,Echevarria:2014xaa,Boglione:2018dqd,Luo:2020hki,Cammarota:2020qcw}. 
At variance with these works, in this paper we start from a recent determination of $f_1$ by our
group~\cite{Bacchetta:2017gcc} and we extract $f_{1T}^\perp$ using the same
formalism, namely for the first time we reconstruct the 
3-dimensional quark density of Eq.~\eqref{e:density} in a fully consistent way within the TMD framework. Later publications have appeared~\cite{Echevarria:2020hpy,Bury:2020vhj,Bury:2021sue} which adopt the same strategy; in the following, we will discuss a comparison with their results. 
%In this work, we start from a recent determination of $f_1$ by our
%group~\cite{Bacchetta:2017gcc} and we extract $f_{1T}^\perp$ using the same
%formalism. Thus, for the first time we consistently reconstruct the full
%3-dimensional quark density of Eq.~\eqref{e:density}. 

Both unpolarized and Sivers TMDs appear in the cross section of polarized
Semi-Inclusive Deep-Inelastic Scattering (SIDIS) and vector-boson production processes. For SIDIS we consider the process $\ell(l) +
N (P) \rightarrow \ell(l') + h(P_h) +X$, where a lepton $\ell$ with momentum
$l$ scatters off a nucleon target $N$ with mass $M$ and momentum $P$.
In the final state, the scattered lepton with momentum $l'=l-q$ is detected,
together with a hadron $h$ with momentum $P_h$ and transverse momentum $P_{hT}$.
We define the usual SIDIS variables $x_{\rm Bj}=Q^2/(2P\cdot q)$, $\; y = P \cdot q/(P \cdot l)$, and $z=P\cdot P_h/(P\cdot q)$.
In this study, we neglect power corrections of order $M^2/Q^2$ and $P_{hT}^2/Q^2$, which allow us also to identify $x_{\rm Bj}=x$.

At leading twist and for a transversely polarized nucleon target $N^\uparrow$, the SIDIS cross section can be parametrized in terms of five structure functions~\cite{Bacchetta:2006tn}:
\begin{eqnarray}
\frac{d\sigma}{dx dy dz d\phi_S d\phi_h dP_{hT}^2} &= &\frac{\alpha^2}{x y Q^2} \, \Bigg\{ A(y) \, F_{UU,T} + B(y) \, \cos 2\phi_h \, F_{UU}^{\cos 2\phi_h} \nonumber \\
& &\hspace{-2.5cm} + |\bm{S}_T| \, \Big[ A(y) \, \sin (\phi_h - \phi_S) \, F_{UT,T}^{\sin (\phi_h - \phi_S)} + B(y) \, \sin (\phi_h + \phi_S) \, F_{UT}^{\sin (\phi_h + \phi_S)} + B(y) \, \sin (3\phi_h - \phi_S) \, F_{UT}^{\sin (3\phi_h - \phi_S)} \Big] \Bigg\}\; ,
\label{eq:xsect}
\end{eqnarray}
where $\alpha$ is the fine structure constant, $\phi_h$ and $\phi_S$ indicate the azimuthal orientations of $\bm{P}_{hT}$ and the target polarization $\bm{S}_T$ in the transverse plane, respectively, the structure functions depend only on $(x, z, P_{hT}^2, Q^2)$, and 
\begin{equation}
A(y) = 1 - y + \frac{1}{2} y^2  \quad , \quad B(y) = 1-y \; .
\end{equation}
The structure function $F_{UU,T}$ can be obtained from the unpolarized cross section after integrating upon all azimuthal angles. The polarized structure function $F_{UT,T}^{\sin (\phi_h - \phi_S)}$ is experimentally measurable through the single spin asymmetry (SSA)
\begin{equation} 
A_{UT}^{\sin(\phi_h-\phi_S)} (x, z, P_{hT}^2, Q^2) = \frac{\int d\phi_S d\phi_h [d\sigma^\uparrow-d\sigma^\downarrow]\sin(\phi_h-\phi_S)}{\int d\phi_S d\phi_h [d\sigma^\uparrow+d\sigma^\downarrow]} \approx
\frac{F_{UT,T}^{\sin(\phi_h-\phi_S)}}{F_{UU,T}}\; . 
\label{e:AUT} 
\end{equation}

Factorization theorems make it possible to write the structure
functions at small transverse momentum ($P_{hT}^2 \ll Q^2$) in
terms of TMDs and to derive their evolution equations. The latter ones are more involved than in the collinear framework because TMDs generally depend on two scales, $\mu^2$ and $\zeta$, that renormalize ultraviolet and rapidity divergences, respectively~\cite{Collins:2011zzd}. These two scales 
%must satisfy the condition $\mu^2 \zeta = Q^4$; they 
are usually chosen to be equal to the virtual photon mass: $\mu^2 = \zeta = Q^2$. 

%The SIDIS cross section can be written in terms of structure
%functions~\cite{Bacchetta:2006tn} that can be measured
%experimentally. Factorization theorems make it possible to write the structure
%functions at small transverse momentum ($P_{hT}^2 \ll Q^2$) in
%terms of TMDs and to derive evolution equations that predict how TMDs change
%as functions of two scales: $\mu^2$ for renormalization and $\zeta$ for rapidity.~\cite{Collins:2011zzd}.
%These two scales are usually chosen to be equal to the virtual photon mass
%$Q^2$. 

%The TMD distributions can be determined from the polarized single spin asymmetry (SSA):
%\begin{equation} 
%A_{UT}^{\sin(\phi_h-\phi_S)} (x, z, \bm{P}_{hT}^2, Q^2) 2\frac{\int d\phi_S %d\phi_h [d\sigma^\uparrow-d\sigma^\downarrow]\sin(\phi_h-\phi_S)}{\int d\phi_S %d\phi_h [d\sigma^\uparrow-d\sigma^\downarrow]} \approx
%\frac{F_{UT,T}^{\sin(\phi_h-\phi_S)}}{F_{UU,T}}\, 
%\label{e:AUT} 
%\end{equation}

%where $\phi_h$ and $\phi_S$ indicate the azimuthal orientations of $\bm{P}_{hT}$ and the target polarization $\bm{S}_T$ in the transverse plane, respectively.
The unpolarized TMD $f_1$ enters the structure function
$F_{UU,T}$, while the Sivers TMD $f_{1T}^\perp$ enters the structure function $F_{UT,T}^{\sin (\phi_h - \phi_S)}$.
%, which occurs in the polarized part of the cross section  weighted by $\sin (\phi_h - \phi_S)$.
Both structure functions can be defined as convolutions of TMDs upon quark
transverse momenta~\cite{Bacchetta:2006tn}, or as Fourier transforms of a product of functions in
$b_T$~\cite{Boer:2011xd}. At leading order in the strong coupling $\alpha_s$ (LO), they read
%The structure functions
%$F_{UU,T}$ and $F_{UT,T}^{\sin(\phi_h-\phi_S)}$
%of Eqs.~\eqref{e:FUU} and \eqref{e:FUT}
%can be written more explicitly as~\cite{Boer:2011xd} 
\begin{align}
\begin{split} 
  F_{UU,T} (x,z,P_{hT}^2, Q^2) &= 
\sum_a e_a^2 x  \int d^2 \bm{k}_T\,  d^2 \bm{P}_T^{}
\, \delta^{(2)}\bigl(z \bm{k}_T + \bm{P}_T^{} - \bm{P}_{h T} \bigr)
f_1^a (x,k_T^2; Q^2) D_1^{a\to h}(z,P_T^2; Q^2)
  \\
%  &=
%\frac{1}{2 \pi} \int^\infty_0 d \bT \bT J_0 (\bT P_{hT}/z) \widetilde{F}_{UU,T} (x,z,\bT^2, Q^2)
%  \\
& = \frac{1}{2\pi} \sum_a e^2_a x \int^\infty_0 d \bT \bT J_0 (\bT P_{hT}/z) \widetilde{f}_1^a (x,\bT^2; Q^2) \widetilde{D}_1^{a\to h}
(z,\bT^2; Q^2)\,,
\end{split}
\\
\begin{split} 
  F_{UT,T}^{\sin(\phi_h-\phi_S)}(x,z,P_{hT}^2, Q^2)
  &= 
- \sum_a e_a^2 x  \int d^2 \bm{k}_T\,  d^2 \bm{P}_T^{}
\, \delta^{(2)}\bigl(z \bm{k}_T + \bm{P}_T^{} - \bm{P}_{h T} \bigr)
\frac{\bm{P}_{h T}\cdot\bm{k}_T}{|\bm{P}_{h T}| M} f_{1T}^{\perp a} (x,k_T^2; Q^2) D_1^{a\to h}(z,P_T^2; Q^2)
\\
%  &= \frac{1}{2 \pi} \int^\infty_0 d \bT
%  \bT^2 J_1(\bT P_{hT}/z) \widetilde{F}_{UT,T}^{\sin(\phi_h-\phi_S)}(x,z,\bT^2, Q^2) 
%  \\
  &= -\frac{M}{2 \pi} \sum_a  e^2_a x \int^\infty_0 d \bT
  \bT^2 J_1(\bT P_{hT}/z) \widetilde{f}_{1T}^{\perp(1)a} (x, \bT ; Q^2)
\widetilde{D}_1^{a\to h} (z,\bT^2; Q^2)\, ,
\label{eq:FUT}
\end{split} 
\end{align}
where $\widetilde{D}_1^{a\to h}$ is the Fourier-transformed expression of the corresponding TMD fragmentation function that describes how the parton $a$ converts into a hadron $h$ with transverse momentum $\bm{P}_{hT}$ and carrying a fraction $z$ of the parton energy. The Fourier transform of the unpolarized TMD is defined as
\begin{equation} 
\widetilde{f}_1^a (x,\bT ^2; Q^2) = \int d^2\kperp e^{i\bm{\bb}_T \cdot
  \kperp} f_1^a (x, k_T^2; Q^2)
= \pi \int_0^\infty dk_T^2 J_0 (\bT k_T) f_1^a (x,k_T^2 ; Q^2)\; , 
\label{eq:f1PCdistr}
\end{equation}
where $J_l$ is the spherical Bessel functions of order $l$. Note that there is a factor $2 \pi$ difference compared to the definition in the extraction of Ref.~\cite{Bacchetta:2017gcc}, denoted as Pavia17, which has been taken into account in the rest of the article. A similar definition holds for $\widetilde{D}_1^{a\to h}$. 

%\\
%\widetilde{f}_{1T}^{\perp (1) a} (x,\bT ^2; Q^2) &= \int d^2\kperp e^{i \bm{\bb}_T \cdot
%  \kperp} \frac{|\kperp|^2}{2 M^2} f_1^a (x,k_T^2; Q^2)
%\label{eq:f1TPCdistr2}
%= \frac{\pi}{M^2}\int_0^\infty d|\kperp|^2 \frac{|\kperp |}{\bT} J_1 (\bT %|\kperp|)
%f_{1T}^{\perp a} (x,k_T^2 ; Q^2)\, ,
%\end{align}  

In Eq.~\eqref{eq:FUT}, we have also introduced the first derivative of the Sivers function in Fourier space~\cite{Boer:2011xd}: 
 \begin{equation}
   \begin{split} 
 &\widetilde{f}_{1T}^{\perp(1)a} (x,\bT^2;Q^2) =   -\frac{2}{M^2}  \partial_ {\bT^2}   \widetilde{f}_{1T}^{\perp a} (x,\bT^2;Q^2) = 
\frac{\pi}{M^2}\int_0^\infty dk_T^2 \frac{k_T}{\bT} J_1 (\bT k_T)
f_{1T}^{\perp a} (x,k_T^2 ; Q^2)\, .
 \label{eq:f1TPCdistr}
%     \\
%   &\frac{n!}{(M^2)^n} \int^\infty_0 \frac{d|\kperp|^2} {2}  \left( \frac{|\kperp|}{\bT}   \right)^n J_n(\bT |\kperp|  ) f_{1T}^{\perp a}  (x,k_\perp^2;Q^2).
   \end{split}
 \end{equation}
 The limit of this formula for $b_T \to 0$ corresponds to the definition of the first $k_T$-moment of the Sivers function:
  \begin{equation}
\lim_{b_T\to 0} \widetilde{f}_{1T}^{\perp(1) a} (x,b_T^2;Q^2) = \int d^2 \bm{k}_T \frac{k_T^2}{2 M^2} f_{1T}^{\perp a} (x,k_T^2; Q^2) = f_{1T}^{\perp(1) a} (x;Q^2) \, ,
\label{e:SivMoments}
\end{equation}
which is an $x$-dependent function and is related to the so-called Qiu-Sterman
function~\cite{Qiu:1998ia,Boer:2003cm}. The precise connection with the
Qiu-Sterman function is nontrivial when considering higher-order corrections
(see, e.g., \cite{Ji:2006ub,Scimemi:2019gge,Ebert:2022cku}). However, these
differences are relevant beyond the order we consider in our analysis.

The unpolarized TMD $f_1$ and the Sivers TMD $f_{1T}^{\perp}$ appear also in the process $A^\uparrow(P_A,S_{AT})+B(P_B) \to \gamma^*/W^\pm/Z^0 + X$, where a hadron $A$ with momentum $P_A$ and transverse polarization $S_{AT}$ scatters off an unpolarized hadron $B$ with momentum $P_B$, producing 
%through annihilation 
a vector boson with four-momentum $q$ and rapidity $y = \frac{1}{2} \log [(q^0 + q_z)/(q^0-q_z)]$, where $\bm{P}_A$ points towards the $\hat{z}$ direction~\cite{Bacchetta:2017gcc}. 

At leading twist and for $q_T \ll q$, the cross section can be parametrized in terms of five structure functions~\cite{Arnold:2008kf}. The relevant terms for the Sivers effect can be expressed as~\cite{Bacchetta:2017gcc,Echevarria:2020hpy}
\begin{equation}
\frac{d\sigma}{dQ^2 dy dq_T^2} = \sigma_0^V \, \Big[ F_{UU}^1 + \sin (\phi_q - \phi_S) \, F_{TU}^1 \Big] \; ,
\label{eq:DYxsect}
\end{equation}
where $Q^2 = q^2$ is the invariant mass of the final state, $\phi_q$ and $\phi_S$ indicate the azimuthal orientations of $\bm{q}_T$ and $\bm{S}_{AT}$ in the transverse plane, respectively, and for $V=\gamma^*, W^\pm, Z^0$ we have
\begin{equation}
\sigma_0^{\gamma^*} = \frac{4\pi^2 \alpha^2}{3 Q^2 s N_c} \; , \quad \sigma_0^{W^\pm} = \frac{\sqrt{2} \pi G_F M_W^2 B_R^W}{s N_c} \, \delta (Q^2 - M_W^2) \; , \quad \sigma_0^{Z^0} = \frac{\sqrt{2} \pi G_F M_Z^2 B_R^Z}{s N_c} \, \delta (Q^2-M_Z^2) \; , 
\label{eq:DYphasespace}
\end{equation}
where $s=(P_A+P_B)^2$, $N_c$ is the number of colors, $G_F$ is the Fermi weak coupling constant, and $B_R^{W/Z}$ is the branching ratio for the decay of vector bosons $W^\pm$ and $Z^0$ with mass $M_W$ and $M_Z$, respectively~\cite{ParticleDataGroup:2020ssz}. 

Again, the structure function $F_{TU}^1$ for the Sivers effect is measurable through the SSA
\begin{equation}
A_N^V (x_A, x_B, q_T^2, Q^2) = \frac{F_{TU}^1}{F_{UU}^1} \; , 
\label{eq:DYSSA}
\end{equation}
where $x_A = e^y Q/\sqrt{s}, \, x_B = e^{-y} Q/\sqrt{s}$, and at LO the structure functions read
\begin{align}
    F_{UU}^1 (x_A, x_B, q_T^2, Q^2) &= \overline{\sum}_{a,a'} |V_{a a'}^V|^2
   \int \frac{d\bT \bT}{2\pi} J_0(\bT q_T) \widetilde{f}_{1}^a (x_A,\bT^2;Q^2) \widetilde{f}_1^{a'} (x_B,\bT^2;Q^2) \; , \\
   F_{TU}^1 (x_A, x_B, q_T^2, Q^2) &= -M \overline{\sum}_{a,a'} |V_{a a'}^V|^2
    \int \frac{d\bT \bT^2}{2\pi} J_1(\bT q_T) \widetilde{f}_{1T}^{\perp(1) a} (x_A,\bT^2;Q^2) \widetilde{f}_1^{a'} (x_B,\bT^2;Q^2) \; ,
\label{eq:DYFF}
\end{align}
where the symbol $\overline{\sum}$ implies adding the contribution of the flavor sum with $A \leftrightarrow B$. For $V=W^\pm$, the $|V_{a a'}^W|^2$ are the elements of the CKM matrix and $a, a'$ run over light quark and antiquark flavors corresponding to $W^\pm$ production:
\begin{equation}
    W^{+} \rightarrow u\overline{d},\,u\overline{s},\,c\overline{d},\,c\overline{s} \; , \quad
    W^{-} \rightarrow d\overline{u},\,d\overline{c},\,s\overline{u},\,s\overline{c} \; .
\end{equation}
For $V=\gamma^*, Z^0$, we have~\cite{Bacchetta:2017gcc,Echevarria:2020hpy}
\begin{equation}
|V_{a a'}^{\gamma^*}|^2 = e_a^2 \delta_{a a'} \; , \quad |V_{a a'}^{Z}|^2 = \Big[ (I_{3a}-2 e_a \sin^2 \theta_W)^2 + (I_{3a})^2 \Big] \, \delta_{a a'} \; ,
\end{equation}
where $\theta_W$ is the Weinberg angle, and the weak isospin $I_{3a}=+1/2$ for $a=u,c,t$ and $-1/2$ for $a=d,s,b$. 
%and for the same  $q\overline{q}$ couples with inverted order of $q_1$ and $q_2$.\\

%We can write the transverse SSA for $W$ production as
%\begin{equation}
%    A_N^W = \frac{d\sigma^\uparrow-d\sigma^\downarrow}{d\sigma^\uparrow+d\sigm%a^\downarrow}
%\end{equation}
%where $\sigma^{\uparrow(\downarrow)}$ denotes the cross section for the process $pp^{\uparrow(\downarrow)}\rightarrow WX$.
%the terms of $A_N^W$ can be written in terms of TMDs:
%\begin{align}
%    d\sigma^\uparrow-d\sigma^\downarrow = -M \sigma_0 \sum_{q1,q2} \left| V_{q1,q2}\right|^2 
%    \int d\boldsymbol{k}_{T 1}d\boldsymbol{k}_{T 2}\delta^{(2)}(\boldsymbol{k}_{T 1}+\boldsymbol{k}_{T 2}-\qT)f_{1T}^{\perp(1)}(x_1,\kpp{1})f_1(x_2,\kpp{2})\\
%    d\sigma^\uparrow+d\sigma^\downarrow = \sigma_0 \sum_{q1,q2} \left| V_{q1,q2}\right|^2
%    \int d\boldsymbol{k}_{T 1}d\boldsymbol{k}_{T 2}\delta^{(2)}(\boldsymbol{k}_{T 1}+\boldsymbol{k}_{T 2}-q_T)f_1(x_1,k_{\perp 1})f_1(x_2,k_{\perp 2})\,.
%\end{align}

%For $Z_0$ we have:

%\begin{align}
%    d\sigma^\uparrow-d\sigma^\downarrow = -M \sigma_0 \sum_{q} e_q^2
%    \int d\bT \bT^2 J_1(\bT \left| k_\perp \right|)\tilde{f}_{1T}^{\perp(1)}(x_A,\bT;\mu_b^2)\tilde{f}_1(x_B,\bT;\mu_b^2)\\
%    d\sigma^\uparrow+d\sigma^\downarrow = \sigma_0 \sum_{q} e_q^2
%   \int d\bT \bT J_0(\bT \left| k_\perp \right|)\tilde{f}_{1}(x_A,\bT;\mu_b^2)\tilde{f}_1(x_B,\bT;\mu_b^2)\,.
%\end{align}
%---- DY formalism

In this work, we take the unpolarized functions $f_1$ and $D_1$ from
the Pavia17 extraction~\cite{Bacchetta:2017gcc}. We extract the Sivers function using the very same
approach: it is based on the TMD framework formulated in Ref.~\cite{Collins:2011zzd}, which in turn elaborates on the original work of Collins, Soper, Sterman~\cite{Collins:1984kg} (hence, in the following we refer to it as the CSS approach). 
%based on the work of Collins, Soper, Sterman
%(CSS)~\cite{Collins:1984kg,Collins:2011zzd}.   
The renormalization group evolution of TMDs is encoded in the so-called Sudakov form factor $S$, which contains the contribution of large logarithms. In this work, we perform the resummation of these logarithms at the next-to-leading-logarithmic (NLL) accuracy, as defined in detail in Ref.~\cite{Bacchetta:2019sam}.\footnote{At this accuracy, in the general formula of the Operator Product Expansion the hard functions and the matching coefficients can be neglected.} 
The expression of $S$ greatly simplifies if the starting scale of evolution is chosen as $\mu_b = 2 e^{-\gamma_E} / b_T$~\cite{Collins:2011zzd}, where $\gamma_E$ is the Euler constant. However, at large $b_T$ the TMD evolution runs into a nonperturbative region and becomes unreliable. In the CSS approach, this pathology is cured by the so-called $b_*$-prescription, which amounts to replacing $\mu_b = 2 e^{-\gamma_E} / b_T$ with $\mu_b = 2 e^{-\gamma_E} / b_*(b_T)$, where $b_*$ is an arbitrary function of $b_T$ with appropriate asymptotic conditions~\cite{Collins:2011zzd}. In accordance with the extraction of the unpolarized TMD~\cite{Bacchetta:2017gcc}, in this analysis we adopt the following function
\begin{equation}
b_* (b_T) = b_{\mathrm{bmax}} \, \left( \frac{1-e^{-b_T^4/b_{\mathrm{max}}^4}}{1-e^{-b_T^4/b_{\mathrm{min}}^4}} \right)^{\frac{1}{4}} \; , 
\label{eq:b*}
\end{equation}
where 
\begin{equation}
b_{\mathrm{bmax}} = 2 e^{-\gamma_E} \; \mbox{GeV}^{-1} \; , \quad b_{\mathrm{min}} = 2 e^{-\gamma_E} / Q \; .
\label{eq:ourb*}
\end{equation}
With this choice, at large $b_T$ the function $b_*(b_T)$ saturates to $b_{\mathrm{max}}$, as already suggested by the CSS approach, and the scale $\mu_b$ freezes at 1 GeV. In this way, the perturbative contributions to the TMD smoothly merge into the nonperturbative region, described by a parametric function (see below). At small $b_T$ (large $k_T$), the TMD formalism is not valid and must match onto the fixed-order formalism. The way the matching is implemented is not unique and the TMD contribution can be arbitrarily modified in this region. At variance with the standard CSS approach, in Eq.~\eqref{eq:b*} we modify the high-transverse-momentum behavior of TMDs as $b_*(b_T) \stackrel{b_T\to 0}{\longrightarrow} b_{\mathrm{min}}$, which implies $\mu_b \to Q$ and preserves a meaningful definition of the integrals inside the Sudakov form factor $S$~\cite{Bacchetta:2017gcc}. The latter prescription partially corresponds to modifying the resummed logarithms as in Ref.~\cite{Bozzi:2010xn} (and similarly in Refs.~\cite{Boer:2014tka,Collins:2016hqq}). 

At NLL accuracy, the TMD evolution of the Sivers function from a starting scale $Q_0^2$ to a generic scale $Q^2$ is formally very similar to the unpolarized TMD $f_1$~\cite{Bacchetta:2017gcc}:
\begin{equation}
  \begin{split}
    \widetilde{f}_{1T}^{\perp(1) a} (x, \bT^2; Q^2) &=  
    e^{S (\mu_b^2, Q^2)} \ e^{g_K(\bT) \ln (Q^2 / Q_0^2)}
%    \\  & \times 
    f_{1T}^{\perp (1)a} (x ; \mu_b^2) \  \widetilde{f}_{1T {\rm NP}}^{\perp} (x, \bT^2) \, .
\label{eq:f1tperpreduced}
  \end{split}
\end{equation}
The $g_K(b_T)$ is the above mentioned universal parametric function that describes the nonperturbative evolution. Together with the perturbative Sudakov form factor $S$, it is the same function that drives the evolution of the unpolarized TMD $f_1$ and is taken from the Pavia17 fit~\cite{Bacchetta:2017gcc}. Without this information, it would not be possible to reliably calculate the Sivers function at the experimental scales. At the initial scale $Q = Q_0 = 1$ GeV, we have $b_{\mathrm{max}} \equiv b_{\mathrm{min}}$ and $\mu_b = Q_0$: the exponentials reduce to unity and evolution effects are switched off. The Sivers function has an intrinsic nonperturbative $b_T$-distribution given by the function $\widetilde{f}_{1T {\rm NP}}^{\perp}$, which needs to be determined from experimental data. For perturbative small values of $b_T$, it can be shown that consistently $\widetilde{f}_{1T {\rm NP}}^{\perp} (x, b_T^2) \to 1$~\cite{Bacchetta:2019sam}. Hence, in the limit $b_T \to 0$ from Eq.~\eqref{eq:f1tperpreduced} we recover Eq.~\eqref{e:SivMoments}: in the perturbative regime the TMD function $\widetilde{f}_{1T}^{\perp(1)}$ is indeed matched through the Operator Product Expansion onto a collinear function represented by the first $k_T$-moment of the Sivers function, $f_{1T}^{\perp (1)}$. 

For $f_{1T}^{\perp (1)}$, we apply the same evolution as the collinear parton density $f_1$ using the HOPPET code~\cite{Salam:2008qg}. This is an approximation of the full
evolution~\cite{Kang:2008ey,Braun:2009mi,Vogelsang:2009pj,Zhou:2008mz,Kang:2012em,Schafer:2012ra,Echevarria:2020hpy}. In order to estimate the impact of the collinear evolution, we compared predictions obtained with our assumptions and predictions with no evolution. We found no significant difference in SIDIS kinematics (containing almost all data used in our fit) because of the limited range in $Q^2$ being spanned. For vector boson production, the difference becomes more relevant, but in both cases the theoretical predictions are small compared to the data, which are few and affected by large errors. 
In conclusion, the approximation in the implementation of collinear evolution does not affect the results of our present fit. The situation will certainly change when more and more precise data at high $Q$ will be available.

The Sivers function must satisfy the positivity
bounds~\cite{Bacchetta:1999kz}~\footnote{The full expression of the positivity
  bound involves also the TMD $g_{1T}$, which is barely known at present~\cite{Bhattacharya:2021twu}. Here, we used a relaxed version where the modulus of $g_{1T}$ is set to zero~\cite{Bacchetta:1999kz}.}
\begin{equation}
\left[ \frac{k_T^2}{2 M^2} \, f_{1T}^{\perp} (x,k_T^2) \right]^2 \leq \frac{k_T^2}{4M^2} f_1^2 (x,k_T^2)
\label{eq:positivity}
\end{equation}
for any value of $x$ and $k_T$. This constraint is essential to guarantee that the quark density distribution of Eq.~\eqref{e:density} is positive everywhere. Therefore, it is convenient to parametrize the nonperturbative function $\widetilde{f}_{1T {\rm NP}}^{\perp}$ of Eq.~\eqref{eq:f1tperpreduced} in momentum space. At the initial scale ($Q_0=1$ GeV), we write the Sivers function 
as 
%a product of a suitably normalized
%$k_T$-dependent function and the first transverse moment,
%$f_{1T}^{\perp (1)}$. The function reads
\begin{equation}
    f_{1T}^{\perp a}(x, k_T^2; Q_0^2) = f_{1T}^{\perp (1) a} (x; Q_0^2) \,   f_{1T\mathrm{NP}}^\perp (x, k_T^2) \, .
    \label{e:guessQ0}
\end{equation}
The nonperturbative term $f_{1T\mathrm{NP}}^\perp$ is given by
\begin{equation}
f_{1T\mathrm{NP}}^\perp (x, k_T^2) = \frac{(1+\lambda_S \,
  k_T^2) \, e^{-k_T^2/M_1^2}}{\pi K(x) \, (M_1^2+\lambda_S M_1^4)} \, 
f_{1 \mathrm{NP}} (x, k_T^2) \; ,
\label{eq:NPparam}
\end{equation}
where $f_{1 \mathrm{NP}}$ is the corresponding nonperturbative term of the unpolarized TMD $f_1$, and is consistently taken from the Pavia17
extraction~\cite{Bacchetta:2017gcc}. More details about the explicit form of the involved functions can be found in~\ref{s:fit}. The $M_1, \, \lambda_S$ are free parameters, and $K (x)$ is a normalization factor to guarantee that the weighted integral of $f_{1T {\rm NP}}^\perp$ is 1 and the proper definition of first $k_T$-moment of the Sivers function is recovered in Eq.~\eqref{e:guessQ0} (see~\ref{s:fit} for details).

The first transverse moment $f_{1T}^{\perp (1) a}$ is parametrized as 
\begin{equation}
  \begin{split} 
    f_{1T}^{\perp (1) a} &(x; Q_0^2) =
    %\frac{1}{F_{\mathrm{max}}} \,
    \frac{N_{\mathrm{Siv}}^a}{G_{\mathrm{max}}^a} \,K(x)\, x^{\alpha_a} (1-x)^{\beta_a}
% \\ &\times
\Big[ 1+A_a \, T_1 (x) + B_a \, T_2 (x) \Big] \, f_1^a(x; Q_0^2) \, ,
\label{eq:f1T1xdep}
\end{split} 
\end{equation}
where $T_n (x)$ are Chebyshev polynomials of order $n$. The unpolarized collinear parton densities $f_1^a$ are taken from the GJR parametrization~\cite{Gluck:2007ck}, consistently with the Pavia17 fit.
%The factors $F_{\mathrm{max}}$ and
The flavor-dependent factor $G_{\mathrm{max}}^a$, and the constraint $|N_{\mathrm{Siv}}^a| \leq 1$, are introduced to guarantee that the Sivers function of Eq.~\eqref{e:guessQ0} satisfies the positivity condition of Eq.~\eqref{eq:positivity} (see~\ref{s:fit} for more details). 
%is introduced to guarantee the positivity bound of the Sivers function of
%Eq.~\eqref{e:guessQ0}~\cite{Bacchetta:1999kz}~\footnote{The full expression of the positivity bound involves also the TMD $g_{1T}$, which is unknown at present. Here, we used a relaxed version where the modulus of $g_{1T}$ is set to zero~\cite{Bacchetta:1999kz}.}
%\begin{equation}
%\left(f_{1T}^{\perp(1)} (x,k_T^2) \right)^2 \leq \frac{k_T^2}{4M^2} f_1^2 (x,k_T^2)
%\end{equation}
%for any value of $x$ and $k_T$. This constraint is essential to guarantee that the quark density distribution of Eq.~\eqref{e:density} is positive everywhere.
The free parameters $N_{\mathrm{Siv}}$, 
%(varying only between $-$1 and 1),
$\alpha, \beta, A, B$ are different for up, down, and sea quarks.

The actual total number of free parameters is 17. We fix them by fitting experimental data for the single transverse-spin asymmetries $A_{UT}^{\sin(\phi_h-\phi_S)}$ of Eq.~\eqref{e:AUT} for SIDIS measurements, and $A_N^V$ of Eq.~\eqref{eq:DYSSA} for vector-boson-production measurements. 
%An accurate extraction requires the inclusion of asymmetry measurements taken by different experimental
%collaborations, covering different ranges of kinematic variables, using
%different type of targets and final-state hadrons.  
In our fit, we include SIDIS measurements by the {\tt HERMES}~\cite{Airapetian:2009ae}, {\tt COMPASS}~\cite{Alekseev:2008aa, Adolph:2016dvl} and JLab collaborations~\cite{Qian:2011py}, and $W/Z$-production measurements taken by the STAR collaboration~\cite{STAR:2015vmv}.
Usually, the SIDIS asymmetries are presented as projections of the same dataset in $x$, $z$, and $P_{hT}$. To avoid fully correlated measurements, we fit  only the $x$ projection because it has a direct impact on the $x$-dependence of the collinear function $f_{1T}^{\perp (1)}$. Similarly, for the STAR dataset we include only one of the projections of the measurements, specifically the data projected in rapidity. 
%Fit results for DY measurements are reported in Tab.~\ref{t:Siv_DY_exp_chi2}, including the projection in $p_T$.
%, since we are mainly interested in the $x$-dependence of the Sivers function.
We select data by applying the same criteria used in the Pavia17 fit for unpolarized TMD, i.e., $Q^2 > 1.4$ GeV$^2$, $0.20<z<0.74$ and
$P_{hT}<\min[0.2Q,\,0.7Qz]+0.5$ GeV~\cite{Bacchetta:2017gcc}. With these kinematic cuts, we have a total of $125$ data points: $30$ from {\tt HERMES}, $82$ from {\tt COMPASS} (32 from the 2009 analysis, and 50 from the 2017 analysis), $6$ from JLab, and $7$ from STAR.  

Similarly to our previous Pavia17 extraction and to other studies of parton densities~\cite{Forte:2002fg,Ball:2010de,Radici:2015mwa},
we perform the fit using the bootstrap method. 
The method consists in creating $\mathcal{M}$ different replicas of the $n$ original data by randomly shifting them with a Gaussian noise with the
same variance as the experimental measurement. Each replica represents the possible outcome of an independent measurement. 
%We then fit each replica separately and we obtain a vector of $\mathcal{M}$ results for each free parameter.
The number $\mathcal{M}$ is fixed by accurately reproducing the mean and standard deviation of the original data points. In our case, it turns out $\mathcal{M}= 200$, which is also consistent with our Pavia17 fit~\cite{Bacchetta:2017gcc}.
%section that was in App A
We denote the replicated measurements as $A_r^{\rm Siv}$, with the $r$ index running from 1 to ${\cal M}$, and with $A_{\mbox{\tiny th}}^{\text{Siv}}\bigl(\{p_r\}\bigr)$ the outcome of the calculated asymmetry using our functional form with the set of parameters $\{p_r\}$. 
Once replicas are generated, a minimization procedure is applied to each replica separately to search for the parameter values, $\{p_{r0}\}$,
that minimize the error function 
\begin{equation}
E_r^2\bigl(\{p_r\}\bigr)= \sum_{i,j=1}^n  \Bigl(A_{r}^{\text{Siv}} - A_{\mbox{\tiny th}}^{\text{Siv}}\bigl(\{p_r\}\bigr)\Bigr)_i 
V_{ij}^{-1}  
\Bigl(A_{r}^{\text{Siv}} - A_{\mbox{\tiny th}}^{\text{Siv}}\bigl(\{p_r\}\bigr)\Bigr)_j \  , 
\label{e:MC_chi2}
\end{equation}
where the covariance matrix is constructed as
\begin{equation}
V_{ij} = \left[  \bigl(\Delta A_{\mbox{\tiny stat}}^{\text{Siv}}\bigr)_i^2 +	\bigl(\Delta A_{\mbox{\tiny sys,unc}}^{\text{Siv}}\bigr)_i^2  + 
\bigl(\Delta A_{\mbox{\tiny th}}^{\text{Siv}}\bigr)_i^2                \right] 
\delta_{ij} +
\sum_{i,j=1}^n \bigl(\Delta A_{\mbox{\tiny corr}}^{\text{Siv}}\bigr)_i \bigl(\Delta A_{\mbox{\tiny corr}}^{\text{Siv}}\bigr)_j \  .
\label{e:covmat}
\end{equation}
For each pair of experimental points $(i, j),$ 
%with $i,j = 1,\ldots, n$, 
the covariance matrix contains the contributions of the statistical $\Delta A_{\mbox{\tiny stat}}^{\text{Siv}}$ and uncorrelated systematic $\Delta A_{\mbox{\tiny sys,unc}}^{\text{Siv}}$ experimental errors, the theoretical error $\Delta A_{\mbox{\tiny th}}^{\text{Siv}}$ due to the uncertainty in the unpolarized TMDs, as well as the correlated experimental uncertainties $\Delta A_{\mbox{\tiny corr}}^{\text{Siv}}$ 
%The terms that appear in the covariance matrix are the statistical and systematic experimental errors including correlated uncertainties 
(like, for example, a $7.3\%$ target polarization correlated uncertainty for the {\tt HERMES} data).  
%and the theoretical error due to the uncertainty in the unpolarized TMDs.
Following the procedure outlined in Ref.~\cite{Bacchetta:2019sam}, we apply the iterative $t_0$-prescription~\cite{Ball:2009qv} in order to avoid the D'Agostini bias that would lead to underestimate the predictions. 
%For each replica separately, we identify the minimum and the corresponding values of best-fit parameters. 
The initial parameter values are chosen randomly within reasonable intervals. For each replica, the goodness of the fit is evaluated using the usual $\chi^2$ test, which corresponds to the error function of Eq.~\eqref{e:MC_chi2}, but with the original experimental data instead of the replicated ones.
%-----------
The maximal information about our results is given by the full ensemble of 200 replicas, combined with the corresponding unpolarized TMD replicas. To
report our results in a concise way, we adopt the following choice:
for any result ($\chi^2$
values, parameter values, resulting distribution functions) we quote intervals containing 68\% of the replicas, obtained by excluding the upper 16\% and lower 16\% values.
These intervals correspond to the $1\sigma$ confidence level only if the
observable's values follow a Gaussian distribution, which is not true in general. 
When it is not possible to draw uncertainty bands,
we report the results obtained using replica 105,
which was selected as a representative replica, since its parameters are the closest to the average ones both in the unpolarized and polarized case.

%section from App A part 2
In Tab.~\ref{t:Siv_parameters} we give the value of the parameters obtained
from our fit. For each one, we quote the central 68\% of the 200 replica
values (by quoting the average 
$\pm$ the semi-difference of the upper and lower limits). Parameters of
replica 105, used for the multidimensional plots, are also given.

\begin{table}
\begin{center}
%\resizebox{\textwidth}{!}{
\begin{tabular}{|c|c|c|c|c|c|c|}
 \hline
          &  $M_1$  & $\lambda_S$  &   $\alpha_d$       &  $\alpha_u$  & $\alpha_s$  & \\
 \hline
   \textbf{All replicas}   &  $0.45\pm 0.10$ & $2.45\pm 2.38$ & $1.64\pm 0.87$ &   $0.51\pm 0.38$ & $0.62\pm 0.52$ & \\
   \hline
 \textbf{ Replica 105} & $0.44$ & $2.00$ & $0.94$ & $0.41$ & $0.61$    & \\
 \hline
  \hline
& $\beta_d$    &  $\beta_u$ & $\beta_s$   &   $A_d$  & $A_u$ &   $A_s$    \\
 \hline
   \textbf{All replicas}   &  $6.59\pm 3.41$ & $2.83\pm 2.26$ & $5.23\pm 4.77$ & $3.66\pm 16.28$ & $-2.33\pm 5.43$ & $ 13.67\pm 22.58$  \\
 \hline
 \textbf{ Replica 105} & $10.00$ & $1.66$ & $8.23$ & $-0.78$ & $-0.58$ & $-1.44$    \\
 \hline
  \hline
& $B_d$  & $B_u$ &   $B_s$ & $N_{\rm Siv}^d$  &  $N_{\rm Siv}^u$ &  $N_{\rm Siv}^s$  \\
 \hline
   \textbf{All replicas}   & $3.40\pm 6.78$ & $2.11\pm 3.81$ & $-0.10\pm 5.13$ & $0.00 \pm 1.00$ & $-0.03\pm 0.46$ & $0.24\pm 0.53$ \\
\hline
 \textbf{Replica 105} &  $0.98$ & $1.12$ & $0.92$ & $-1.00$ & $0.42$ & $0.28$     \\
\hline
\end{tabular}
%}
\caption{Values of the best fit parameters for the Sivers distribution. For each parameter, the upper row contains the central 68\% confidence interval obtained from 200 replicas by indicating the average value $\pm$ the semi-difference of the upper and lower limits. The lower row refers to the replica 105 whose parameter values are the closest to the average ones.}
\label{t:Siv_parameters}
\end{center}
\end{table}
%------------------

We obtain an excellent agreement between the experimental measurements
and our theoretical prediction, with an overall value of $\chi^2 / $d.o.f. $= 1.12 \pm 0.04$ (total $\chi^2=121\pm 5$). In~\ref{s:fitplots}, we collected all figures that show the quality of our fit. Our parametrization is able to describe very well the {\tt COMPASS} 2009 data set~\cite{Alekseev:2008aa} (32 points with $\chi^2 = 26.5 \pm 4.2$; see Fig.~\ref{f:Siv_C09}), the {\tt COMPASS} 2017 data set~\cite{Adolph:2016dvl} (50 points with $\chi^2 = 31.3 \pm 3.8$; see Figs.~\ref{f:Siv_C17piu} and \ref{f:Siv_C17meno}), and the JLab data set~\cite{Qian:2011py} (6 points with $\chi^2 = 4.1 \pm 0.7$; see Fig.~\ref{f:Siv_JLab}). The agreement with the {\tt HERMES} data set~\cite{Airapetian:2009ae} is somewhat worse (30 points with $\chi^2 = 46.2 \pm 3.8$; see Fig.~\ref{f:Siv_Hermes}). We checked that the largest contribution to the $\chi^2$ comes from the subset of data with $K^-$ in the final state~\cite{Signori:2013mda}. Looking at the previous figures it is important to notice, as a check of the results validity, that our predictions well describe also the $z$ and $P_{hT}$ distributions, even if those projections of the data were not included in the fit (see~\ref{s:fitplots} for more details). 

The agreement with vector-boson-production {\tt STAR} measurements~\cite{STAR:2015vmv} is worse than the SIDIS case, with a $\chi^2=13.97\pm 0.6$ for $7$ points. However, the lower number of points (see Fig.~\ref{f:Siv_STAR}) indicates that {\tt STAR} data  have less influence on the global fit than the SIDIS data. In any case, we observe that our predictions follow the sign of the measurements, being negative for $W^+$ and positive for $W^-$ and $Z^0$. The agreement is similar for the data points projected in $p_T$ not included in the fit (see~\ref{s:fitplots} for more details). 
%(More information about the fit procedure, the best-fit parameters and the agreement with data can be found in App.~\ref{s:fit}.)

%%%%%%%%%%%%%%%%%% Fig.1 %%%%%%%%%%%%%%%%%
\begin{figure}
  \begin{center}
    \includegraphics[width=\textwidth]{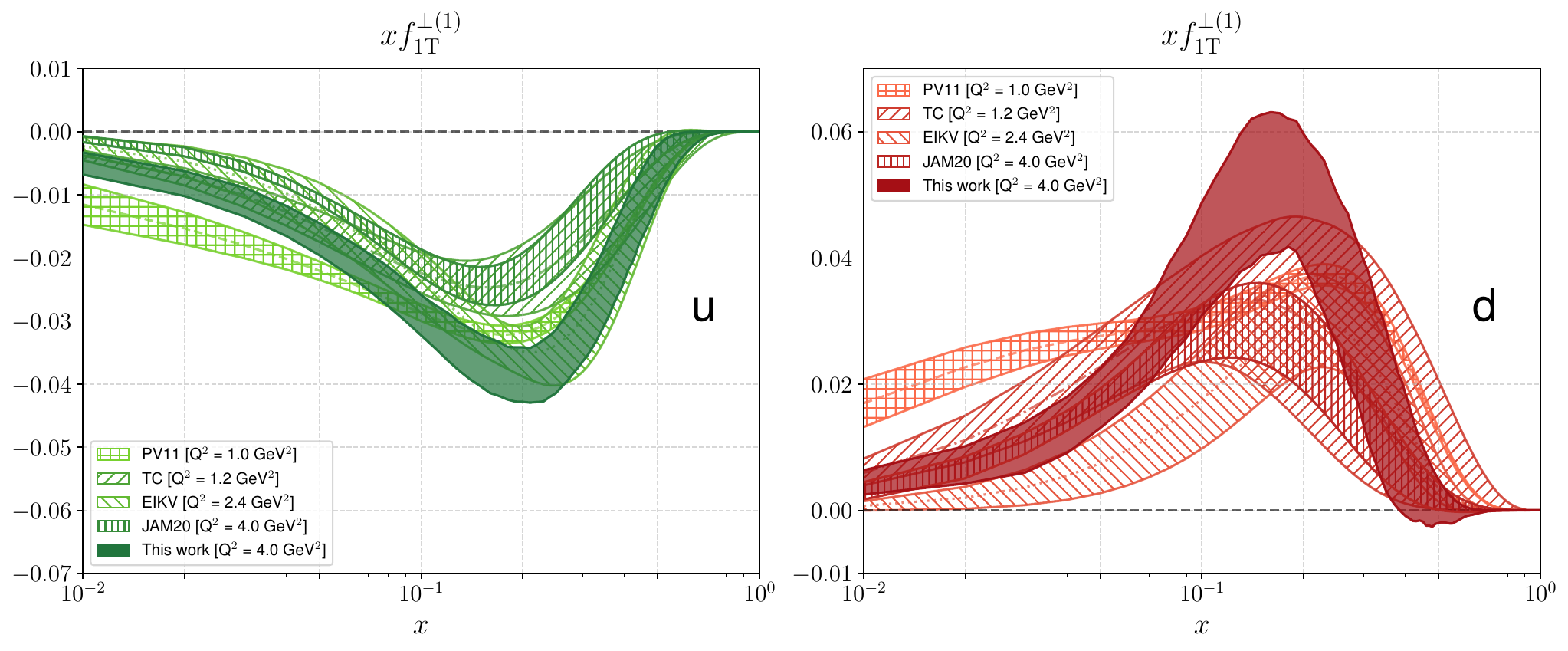}
  \end{center}
\vspace{-0.2cm}
\caption{The first transverse moment $x f_{1T}^{\perp (1)}$ of the Sivers
  TMD as a function of $x$ for the up (left panel) and down quark (right
  panel). Solid band: the 68\% confidence interval obtained in
  this work at $Q^2 = 4$ GeV$^2$. Hatched bands from
  PV11~\cite{Bacchetta:2011gx}, EIKV~\cite{Echevarria:2014xaa}, TC18~\cite{Boglione:2018dqd}, JAM20~\cite{Cammarota:2020qcw} parametrizations, and at different $Q^2$ as indicated in the figure.}
\label{f:Sivers_band}
\end{figure}
%%%%%%%%%%%%%%%%%%%%%%%%%%%%%%%%%%%%%%%

In Fig.~\ref{f:Sivers_band}, we show the first transverse moment
$x f_{1T}^{\perp (1)}$ (Eq.~\eqref{e:SivMoments}, multiplied by $x$)
as a function of $x$ at $Q_0 = 2$ GeV for the up (left panel) and down
quark (right panel).
We compare our results (solid band) with other parametrizations available in
the literature~\cite{Bacchetta:2011gx,Echevarria:2014xaa,Boglione:2018dqd,Cammarota:2020qcw}
(hatched bands, as indicated in the figure).  
In agreement with previous studies, the distribution for the up quark is negative, while for the down quark is positive and both have a similar magnitude. The Sivers function for sea quarks is very small and compatible with zero.

The authors of Ref.~\cite{Echevarria:2020hpy} also find results very similar
to the ones in Fig.~\ref{f:Sivers_band} when they fit the same SIDIS data and
{\tt COMPASS} Drell--Yan data with pion beams~\cite{COMPASS:2017jbv}. In this
case, they also compute predictions for $W^\pm$ and $Z^0$ production at {\tt
  STAR} kinematics which are very close to our fitted bands displayed in
Fig.~\ref{f:Siv_STAR}. Their strategy is very similar to the one adopted in
this work but at higher perturbative accuracy, although their unpolarized TMDs
are not obtained from an actual fit. However, when they include the {\tt STAR}
data in the global fit they artificially increase the statistical weight of
those data by a factor $\sim 13$. Their global $\chi^2$ largely deteriorates
and the uncertainty on the Sivers function significantly increases.
%We believe that this procedure is not correct and introduces unphysical
%tensions among the various data sets.
Our finding is that because of large experimental errors {\tt STAR} data does not affect much our final results when including them in the global fit, as discussed in detail in~\ref{s:fitplots}. 

\begin{figure}[ht]\
  \begin{center}
    \begin{tabular}{ccc}
      \includegraphics[width=0.40\textwidth]{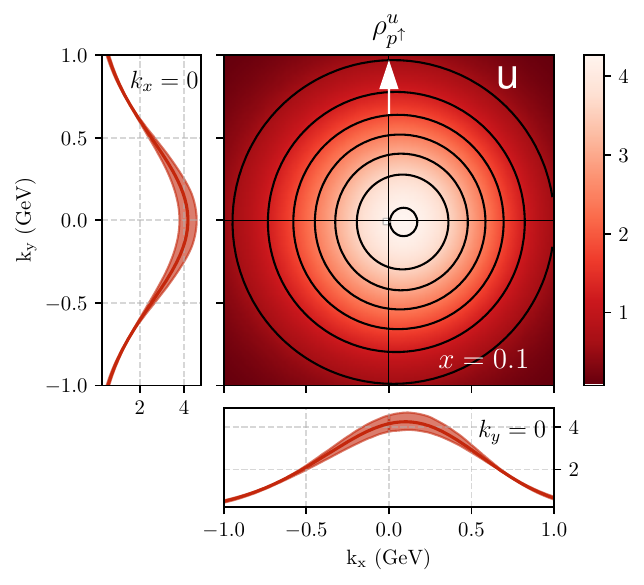}& &
      \includegraphics[width=0.40\textwidth]{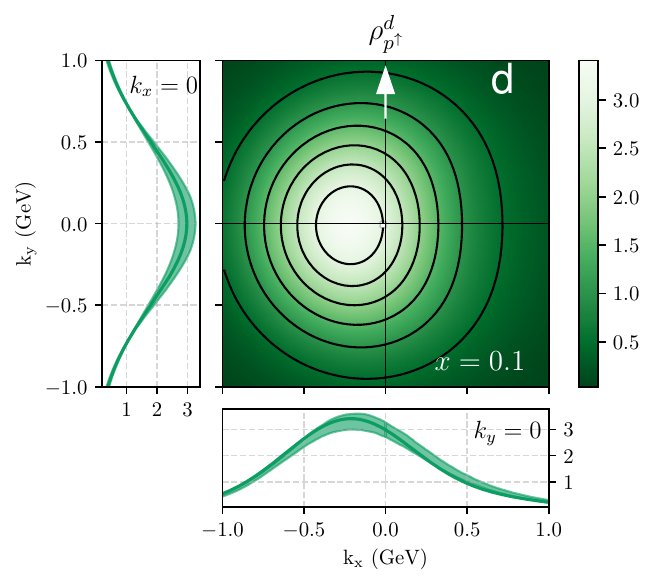}
      \\
      \includegraphics[width=0.40\textwidth]{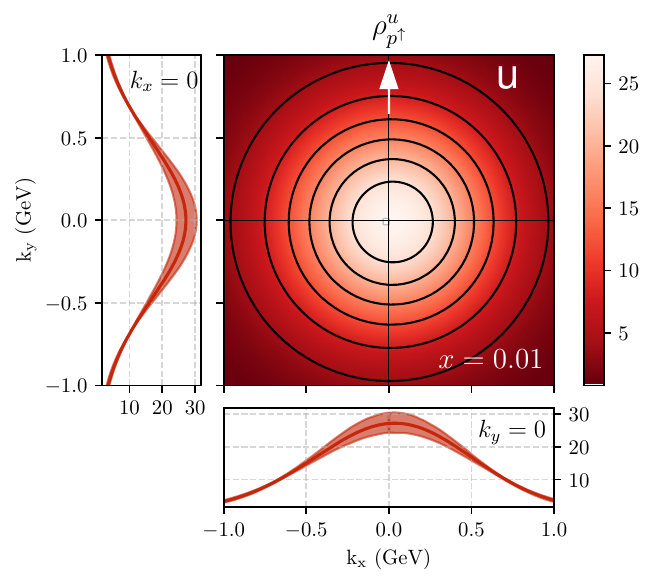}& &
      \includegraphics[width=0.40\textwidth]{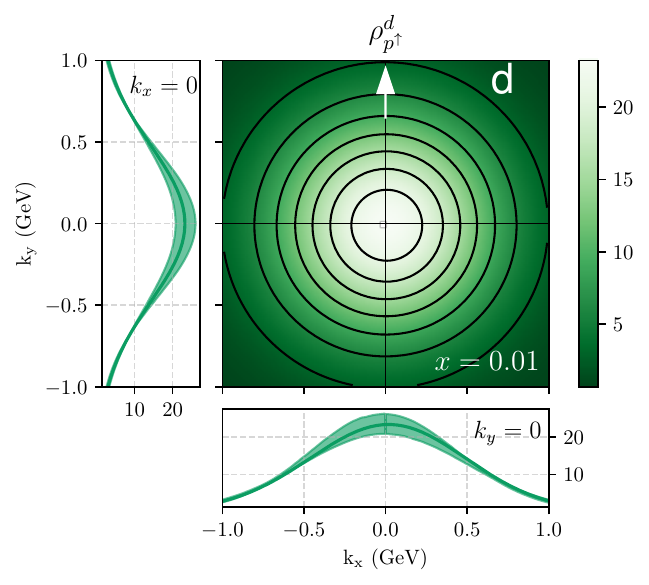}
    \end{tabular}
    \end{center}
\vspace{-0.2cm}   
\caption{The density distribution $\rho_{p^\uparrow}^a$ of an unpolarized quark
  with flavor $a$ in a proton
  polarized along the $+y$ direction and moving towards the
  reader, as a function of $(k_x, k_y)$ at $Q^2 = 4$ GeV$^2$. Left panels for
  the up quark, right panels for the down quark. Upper panels for results at
  $x=0.1$, lower panels at $x=0.01$. For each panel, lower ancillary plots
  represent the 68\% uncertainty band of the distribution at $k_y = 0$ (where
  the effect of the distortion due to the Sivers function is maximal)  while
  left ancillary plots at $k_x = 0$ (where the distribution is the same as for
  an unpolarized proton). Results in the contour plots and the solid lines in
  the projections correspond to replica 105 (see text).} 
\label{f:density}
\end{figure}

The authors of Ref.~\cite{Bury:2021sue} also perform a consistent extraction
of both unpolarized and Sivers TMDs, and build contour plots of the density
distribution in Eq.~\eqref{e:density} similar to Fig.~\ref{f:density}. A
direct comparison is more difficult because the evolution of TMDs is achieved
in a different framework, and the classification of the perturbative accuracy
does not match the standard described in Ref.~\cite{Bacchetta:2019sam}. The
displayed $x$-dependence of their Qiu-Sterman function (or related first
$k_T$-moment of the Sivers function as in Eq.~\eqref{e:SivMoments}) is roughly
similar, at least for up and down quarks. However, the sea-quark channel shows
large oscillations at large $x$, which entail a strong breaking of the
positivity constraint of Eq.~\eqref{eq:positivity}.  

In general, the result of a fit is biased whenever a specific fitting
functional form is chosen at the initial scale. In our case, we tried to
reduce this bias by adopting a flexible functional form, as it is evident
particularly in Eq.~\eqref{eq:f1T1xdep}. Nevertheless, we stress that our
extraction is still affected by this bias  and extrapolations outside the
range where data exist ($0.01 \lesssim x \lesssim 0.3$) should be taken with due care.
At variance with previous studies, in the denominator of the asymmetries in
Eqs.~\eqref{e:AUT} and~\eqref{eq:DYSSA} we are using unpolarized TMDs that were
extracted from data in our previous Pavia17 fit, with their own uncertainties.
Therefore, our uncertainty bands in Fig.~\ref{f:Sivers_band} represent a
realistic estimate of the statistical error of the Sivers function. 

%\begin{widetext}

%%%%%%%%%%%%%%%%%%  Fig. 2 %%%%%%%%%%%%%%%%%%%
%%%%%%%%%%%%%%%%%%%%%%%%%%%%%%%%%%%%%%%%

%\end{widetext}

In Fig.~\ref{f:density}, we show the density distribution $\rho^a_{p^\uparrow}$
of an unpolarized quark $a$ in a transversely polarized proton defined in
Eq.~\eqref{e:density}, at $x=0.1$ (upper panels) and $x=0.01$ (lower
panels) and at the scale $Q^2 = 4$ GeV$^2$.
The proton is moving towards the reader and is polarized along the
$+y$ direction. Since the up Sivers function is negative, the induced
distortion is positive along the $+x$ direction for the up quark (left
panels), and opposite for the down quark (right panels).

At $x=0.1$ the distortion due to the Sivers effect is evident, since we are
close to the maximum value of the function shown in
Fig.~\ref{f:Sivers_band}. The distortion 
%is opposite for up and down quarks, reflecting the opposite sign of the Sivers function. It 
is more pronounced for down quarks, because the Sivers function is larger and at the same time the unpolarized TMD is smaller. The peak positions are approximately $(k_x)_{\text{max}} \approx 0.1$ GeV for up quarks and $-0.15$ GeV for down quarks. At lower values of $x$, the distortion
disappears. These plots suggest that a virtual photon hitting a transversely
polarized proton effectively ``sees'' more up quarks to its right and more
down quarks to its left in momentum space.

%The peak positions are approximately $(k_x)_{\text{max}} \approx 0.1$ GeV for up quarks and $-0.15$ GeV for down quarks. 
%We can use this information to quantify in a simple way the Sivers distortion.
%Since $1/(k_x)_{\mathrm{max}} \approx 1/100 \  \mathrm{MeV}^{-1} \approx 2$ fm, $e_u = 2/3 \, |e| \approx 10^{-19} C$, then 
%\begin{equation}
%    e_u/(k_x)_{\mathrm{max}} \approx 2 \times 10^{-34} \  \mathrm{C} \cdot \mathrm{m} \approx 0.6 \times 10^{-4} \  \mathrm{debye} \, .
%\end{equation}
%To have a feeling of the order of magnitude of this distortion, we can estimate the expression
%$e_q/(k_x)_{\text{max}} \approx 2 \times 10^{-34} {\rm C} \times {\rm m}
%\approx 0.6 \times 10^{-4}$ debye, 
%Just for comparison, this is about $10^{-5}$ times smaller than the electric dipole of a water molecule and about 10 times greater than a nuclear magneton divided by $c$. 

The existence of this distortion requires two ingredients. First of all, the
wavefunction describing quarks inside the proton must have a component with
nonvanishing angular momentum. Secondly, effects due to final state interactions should be
present~\cite{Brodsky:2002cx}, which in Feynman gauge can be described as the exchange of Coulomb
gluons between the quark and the rest of the proton~\cite{Ji:2002aa}.
In simplified models~\cite{Pasquini:2019evu},
it is possible to separate these two ingredients and obtain an
estimate of the angular momentum carried by each quark~\cite{Burkardt:2002ks}.
It turns out that up quarks give almost 50\% contribution to the
proton's spin, while all other quarks and antiquarks give less than
10\%~\cite{Bacchetta:2011gx}. We will leave this model-dependent study to a future publication. 
A model-independent estimate of quark angular momentum requires the determination of parton distributions
that depend simultaneously on momentum and position~\cite{Ji:1996ek,Lorce:2011ni}. Nevertheless, 
the study of TMDs, and of the Sivers function in particular, can provide important constraints on
models of the nucleon~\cite{Burkardt:2015qoa} and test lattice QCD computations~\cite{Yoon:2017qzo}.

%Again, just to have a
%feeling of the order of magnitude of the involved angular momenta, we can take
%a typical value for the average position of the quarks in the proton equal to
%$R\approx0.5 \times 10^{-15}m$ and 
%consider the product
%$R k_{x \text{max}}\approx 0.25 \hbar$. 

In the near future, more data are expected from experiments
at Jefferson Laboratory and CERN. Pioneering measurements in Drell-Yan
processes with pion beams have been already reported~\cite{COMPASS:2017jbv}, but they are not included in the present analysis because 
%of their relatively large uncertainties. 
we do not have yet a consistent description of quark unpolarized TMDs in a pion. 
In the longer term, the recently approved Electron Ion Collider project~\cite{Boer:2011fh,Accardi:2012qut} will provide a large amount of
data in different kinematic regions compared to present experiments~\cite{AbdulKhalek:2021gbh}. With this abundance of data, we will be able to reduce the error bands, extend the range of validity of the extractions to lower and higher values of $x$, and obtain a much more detailed knowledge of the 3-dimensional distribution of partons in momentum space. 

\section*{Acknowledgements}

%\begin{acknowledgments}
%\begin{addendum}
%\item
  This work is
supported by the European Research Council (ERC) under the European
Union's Horizon 2020 research and innovation program (grant agreement
No. 647981, 3DSPIN), and by the European Union’s Horizon 2020 research and innovation program under agreement STRONG - 2020 - No 824093.
This material is based upon work supported by the U.S. Department of Energy, Office of Science, Office of Nuclear Physics under contract DE-AC05-06OR23177.
% \item[Correspondence] Correspondence and requests for materials
%should be addressed to Alessandro Bacchetta~(email: alessandro.bacchetta@unipv.it).
%\end{addendum}
%\end{acknowledgments}

\appendix
  
%\begin{widetext}

\section{Details about the fitting functional form}
\label{s:fit}

The functional form we chose to parametrize the Sivers function is built in order to automatically satisfy the positivity bound
\begin{equation}
\left[ \frac{k_T^2}{2M^2} \, f_{1T}^{\perp} (x,k_T^2) \right]^2 \leq \frac{k_T^2}{4M^2} f_1^2 (x,k_T^2) \; .
\label{eq:Apositivity}
\end{equation}
It is given by
\begin{equation}
    f_{1T}^{\perp a}(x, k_T^2; Q_0^2) = f_{1T}^{\perp (1) a} (x; Q_0^2) \,   f_{1T\mathrm{NP}}^\perp (x, k_T^2) \; ,
\label{e:AguessQ0}
\end{equation}
with
\begin{equation}
f_{1T\mathrm{NP}}^\perp (x, k_T^2) = \frac{(1+\lambda_S \,
  k_T^2) \, e^{-k_T^2/M_1^2}}{\pi K(x) \, (M_1^2+\lambda_S M_1^4)} \, 
f_{1 \mathrm{NP}} (x, k_T^2) \; .
\label{eq:ASivNP}
\end{equation}
The $f_{1 \mathrm{NP}}$ is the nonperturbative part of the unpolarized TMD $f_1$ and is taken from the Pavia17 extraction~\cite{Bacchetta:2017gcc}:
\begin{equation}
    f_{1 \mathrm{NP}} (x, k_T^2) = \frac{1}{\pi} \frac{1+\lambda k_T^2}{g_{1}(x)+\lambda g_{1}^2(x)}e^{-k_T^2/g_{1}(x)},
\end{equation}
with
\begin{align} 
 g_1 (x) &= N_1 \;  
\frac{(1-x)^{\alpha} \  x^{\sigma} }{ (1-\hat{x})^{\alpha} \  \hat{x}^{\sigma} } \, ,\qquad \hat{x}=0.1 \, .
\label{e:kT2_kin}
\end{align}
The distributions of values for the parameters $N_1$, $\alpha$, $\sigma$, $\lambda$ are obtained from the Pavia17 fit and can be found in the NangaParbat repository~\footnote{https://github.com/MapCollaboration/NangaParbat}.
For convenience, we report here the values for the relevant replica 105
\begin{align}
N_1 &= 0.285 \ {\rm GeV}^2,
&
\alpha & = 2.98,
& 
\sigma &= 0.173, 
& 
\lambda &= 0.39 \ {\rm GeV}^{-2} \; .
\end{align}

In Eq.~\eqref{eq:ASivNP}, the $M_1, \, \lambda_S$ are free parameters. The $K(x)$ is a normalization factor to guarantee that the weighted integral of $f_{1T {\rm NP}}^\perp$ is 1 and the proper definition of the first $k_T$-moment of the Sivers function is recovered in Eq.~\eqref{e:AguessQ0}. It is given by
\begin{align}
\label{eq:KSiv}
K (x)&\equiv   \pi \int dk_T^2 \frac{k_T^2}{2M^2} \frac{\Big(1+\lambda_S k_T^2\Big)}{\pi \Big(M_1^2+\lambda_S M_1^4\Big)} e^{-k_T^2/M_1^2} \frac{(1+\lambda k_T^2)}{\pi\Big(g_1(x)+\lambda [g_1(x)]^2 \Big)}e^{-k_T^2/g_1(x)}\\
 &=\,
 \frac{g_1(x) \, M_1^2}{2 \pi M^2 \Big( 1+ \lambda g_1(x)\Big) \,  \Big( g_1(x)+M_1^2 \Big)^2 \, \Big( 1 + \lambda_S M_1^2 \Big)} \, \left[ 1 + 2 (\lambda + \lambda_S) \, \frac{g_1(x) \, M_1^2}{g_1(x) + M_1^2} + 6 \lambda \lambda_S \, \left( \frac{g_1(x) \, M_1^2}{g_1(x) + M_1^2} \right)^2 \right] \; . \nonumber
\end{align}

The Fourier transform of $f_{1T\mathrm{NP}}^\perp$ in Eq.~\eqref{eq:ASivNP} reads 
\begin{align}
\widetilde{f}_{1T {\rm NP}}^{\perp} (x, \bT^2) &= \pi \int dk_T^2 \, J_0 (k_T b_T) \, f_{1T {\rm NP}}^{\perp} (x, k_T^2) = \frac{\exp\left(-\frac{b_T^2}{4} \, \frac{g_1 M_1^2}{g_1+M_1^2}\right) }{\pi \, K(x) \, \Big( 1 + \lambda_S M_1^2 \Big) \, \Big( 1+ \lambda g_1 \Big) \, \Big( g_1 + M_1^2 \Big)} \nonumber \\[2pt]
&\!\!\!\!\!\!\!\! \times \Bigg\{ 1 + \left( \lambda + \lambda_S + 2 \lambda \lambda_S \frac{g_1 M_1^2}{g_1 + M_1^2} \right) \, \frac{g_1 M_1^2}{g_1 + M_1^2} \, \left( 1 - \frac{b_T^2}{4}\, \frac{g_1 M_1^2}{g_1 + M_1^2} \right) - \frac{b_T^2}{2}\, \left( \frac{g_1 M_1^2}{g_1 + M_1^2} \right)^3 \, \lambda \lambda_S \, \left( 1 - \frac{b_T^2}{8}\, \frac{g_1 M_1^2}{g_1 + M_1^2} \right) \, \Bigg\} \nonumber \\[4pt]
&= \frac{2 M^2 \, (g_1 + M_1^2) \, \exp\left(-\frac{b_T^2}{4} \, \frac{g_1 M_1^2}{g_1+M_1^2}\right) }{g_1 M_1^2 \, \left[ 1 + \left( \lambda + \lambda_S + 3 \lambda \lambda_S \, \frac{g_1 M_1^2}{g_1 + M_1^2} \right) \, 2 \, \frac{g_1 M_1^2}{g_1 + M_1^2}  \right]} \nonumber \\[2pt]
&\!\!\!\!\!\!\!\! \times \Bigg\{ 1 + \left( \lambda + \lambda_S + 2 \lambda \lambda_S \frac{g_1 M_1^2}{g_1 + M_1^2} \right) \, \frac{g_1 M_1^2}{g_1 + M_1^2} \, \left( 1 - \frac{b_T^2}{4}\, \frac{g_1 M_1^2}{g_1 + M_1^2} \right) - \frac{b_T^2}{2}\, \left( \frac{g_1 M_1^2}{g_1 + M_1^2} \right)^3 \, \lambda \lambda_S \, \left( 1 - \frac{b_T^2}{8}\, \frac{g_1 M_1^2}{g_1 + M_1^2} \right) \, \Bigg\} \; .
\label{eq:f1t1perpLamS}
\end{align}

The first transverse moment is parametrized as 
\begin{equation}
  \begin{split} 
    f_{1T}^{\perp (1) a} &(x; Q_0^2) =
    %\frac{1}{F_{\mathrm{max}}} \,
    \frac{N_{\mathrm{Siv}}^a}{G_{\mathrm{max}}^a} \,K (x)\, x^{\alpha_a} (1-x)^{\beta_a}
% \\ &\times
\left[1+A_a \, T_1 (x) + B_a \, T_2 (x) \right] \, f_1^a(x; Q_0^2) \, ,
\label{eq:A_f1T1xdep}
\end{split} 
\end{equation}
where $T_n (x)$ are Chebyshev polynomials of order $n$, and the unpolarized collinear parton densities $f_1^a$ are taken from the GJR parametrization~\cite{Gluck:2007ck}, consistently with the Pavia17 fit. 
The flavor-dependent factor $G_{\mathrm{max}}^a$ is defined as
\begin{align}
    G_{\mathrm{max}}^a =
    \max_{k_T} \Biggl[ \frac{k_T}{\pi M} \, \biggl|\frac{1+\lambda_S k_T^2}{M_1^2+\lambda_S M_1^4}\biggr| \, e^{-k_T^2/M_1^2} \Biggr]\; \max_x \Biggl[ x^{\alpha_a} (1-x)^{\beta_a} \, \bigl| 1+A_a \, T_1 (x) + B_a \, T_2 (x) \bigr| \, \Biggr] \; .
\end{align}
Together with the constraint $|N_{\mathrm{Siv}}^a| \leq 1$, it is introduced into Eq.~\eqref{eq:A_f1T1xdep} to guarantee that the Sivers function of Eq.~\eqref{e:AguessQ0} satisfies the positivity condition of Eq.~\eqref{eq:Apositivity}. 
For the relevant replica 105 we have
\begin{align}
    G_{\mathrm{max}}^u & = 1.99 \times 10^{-2},
    &
    G_{\mathrm{max}}^d & = 5.58 \times 10^{-4},
    &
    G_{\mathrm{max}}^s & = 2.42 \times 10^{-3}.
\end{align}

\section{Comparison with data}
\label{s:fitplots}

In this section, we present figures showing the quality of our fit. In all plots, our results are represented by solid bands whenever data points are actually included in the fit, while hatched bands are predictions for the same measurements projected over different kinematic variables. The predictions are obtained by integrating upon the variables over which they are not projected.

\begin{figure}[h!]
\begin{center}
\includegraphics[width=\textwidth]{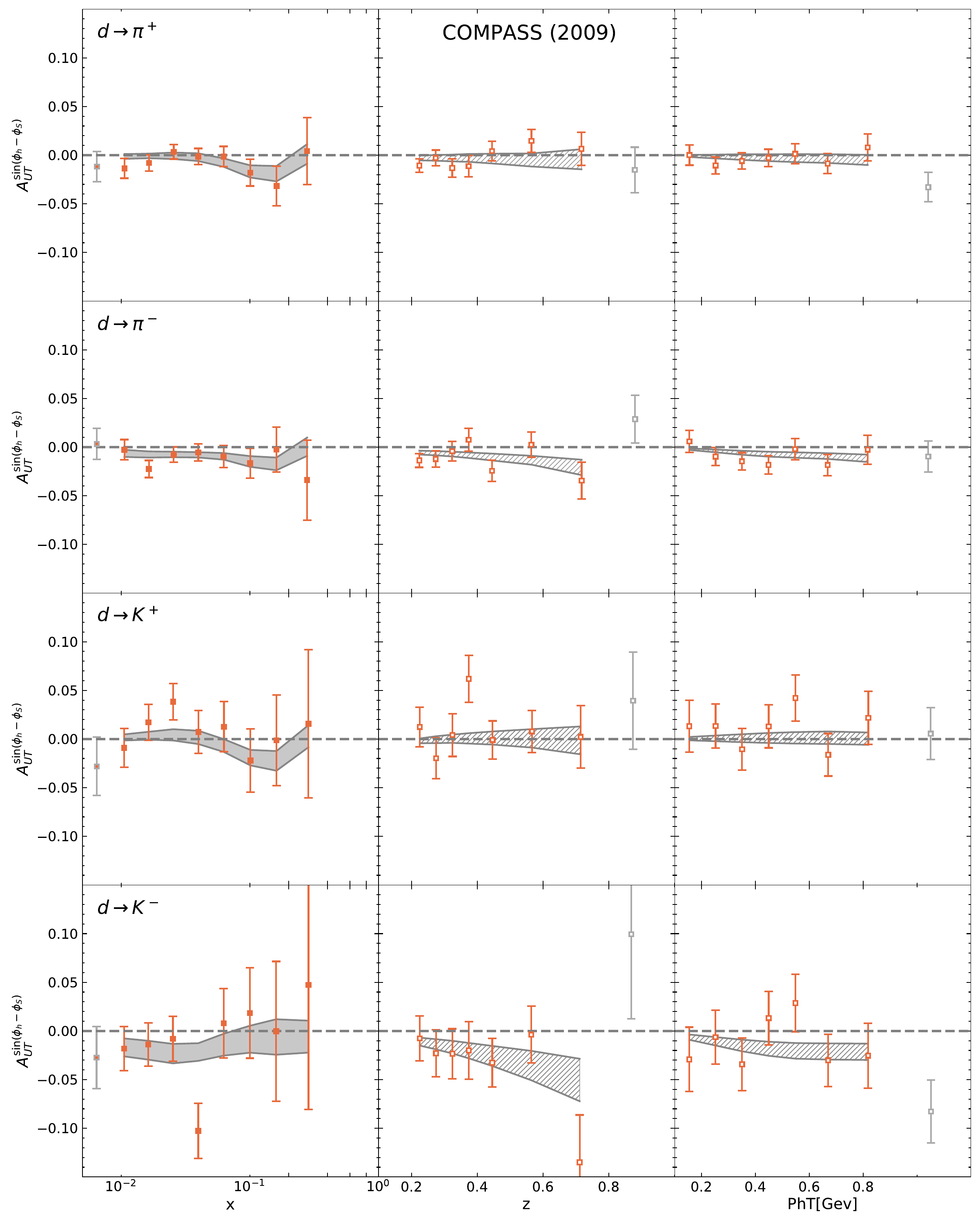}
\end{center}
\caption{COMPASS 2009 Sivers asymmetries from SIDIS off a deuteron target
  ($^6$LiD) with production of $\pi^+$,  $\pi^-$, $K^+$, $K^-$ in the final
  state~\cite{Alekseev:2008aa}, presented as function of $x$, $z$, $P_{hT}$. Only the $x$-dependent projections have been included in the fit (solid bands), the dependence on other variables is predicted (hatched bands). } 
\label{f:Siv_C09}
\end{figure}

\begin{figure}[h!]
\begin{center}
\includegraphics[width=\textwidth]{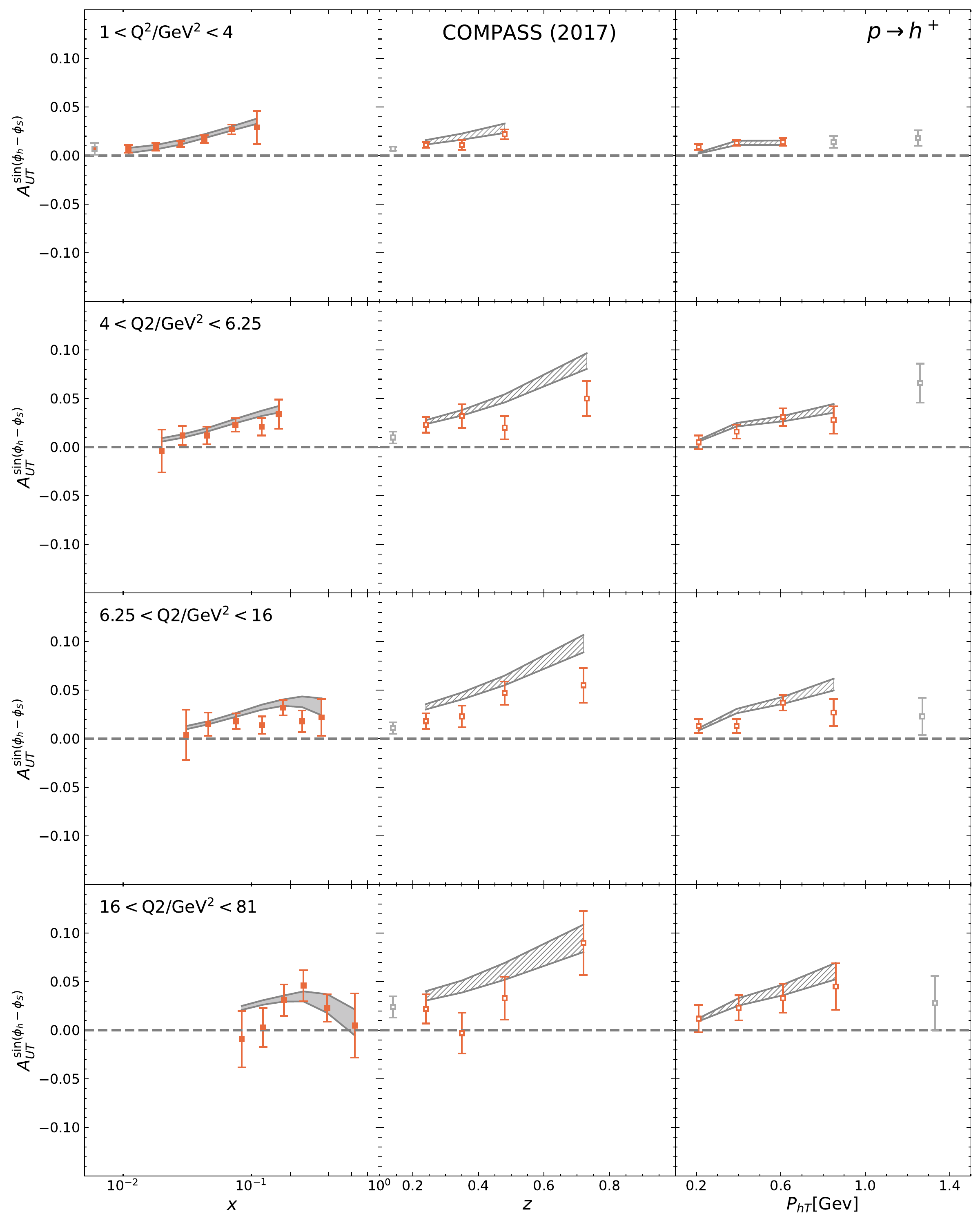}
\end{center}
\caption{COMPASS 2017 Sivers asymmetries from SIDIS off a proton target (NH$_3$) with production of positive hadrons $h^+$~\cite{Adolph:2016dvl}, presented as function of $x$, $z$, $P_{hT}$ and divided in four different $Q^2$ bins. Same notation as in previous figure.} 
\label{f:Siv_C17piu}
\end{figure}

\begin{figure}[h!]
\begin{center}
\includegraphics[width=\textwidth]{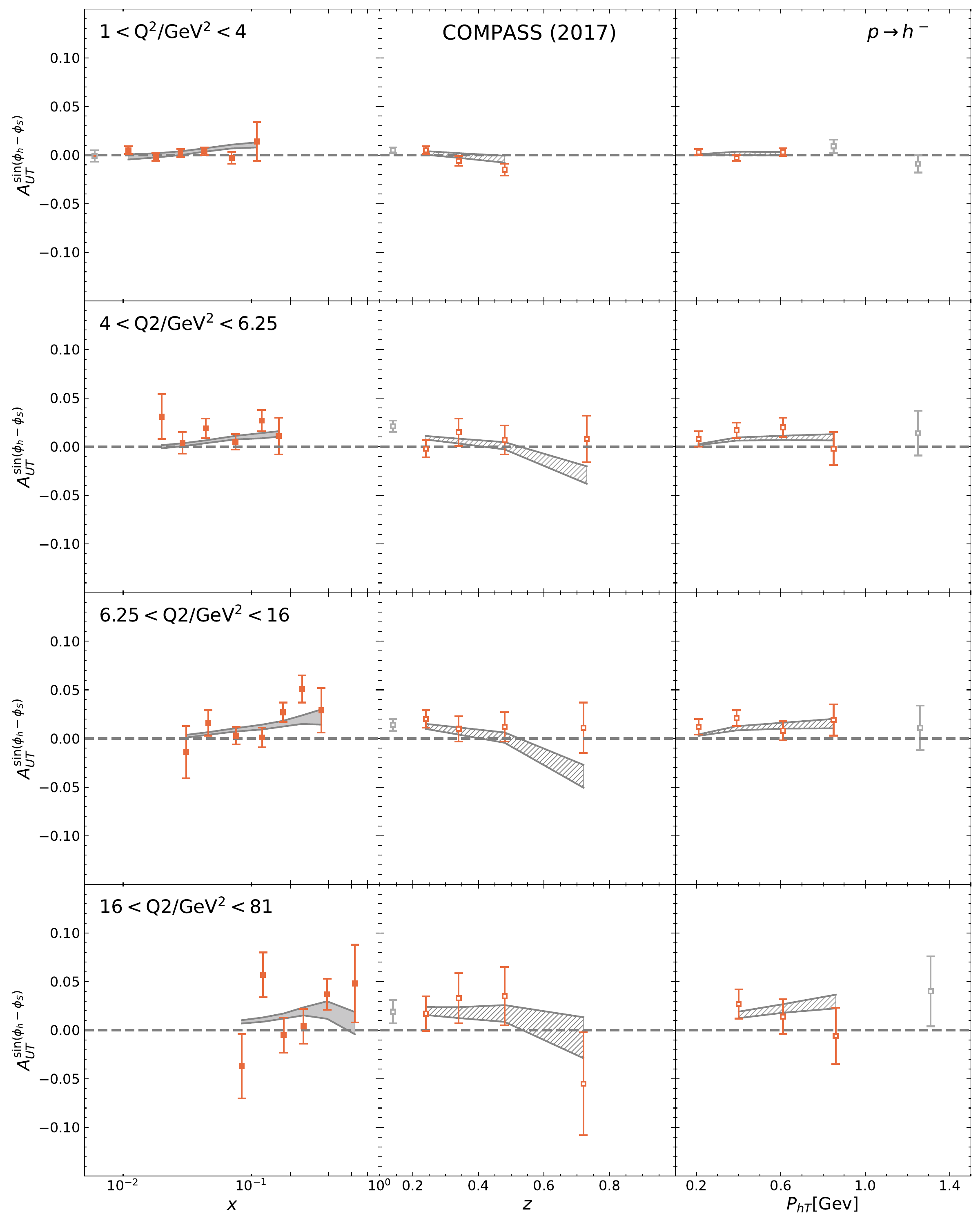}
\end{center}
\caption{COMPASS 2017 Sivers asymmetries from SIDIS off a proton target (NH$_3$) with production of negative hadrons $h^-$~\cite{Adolph:2016dvl}, presented as function of $x$, $z$, $P_{hT}$ and divided in four different $Q^2$ bins. Same notation as in Fig.~\ref{f:Siv_C09}.} 
\label{f:Siv_C17meno}
\end{figure}

\begin{figure}[h!]
\begin{center}
\includegraphics[width=0.7\textwidth]{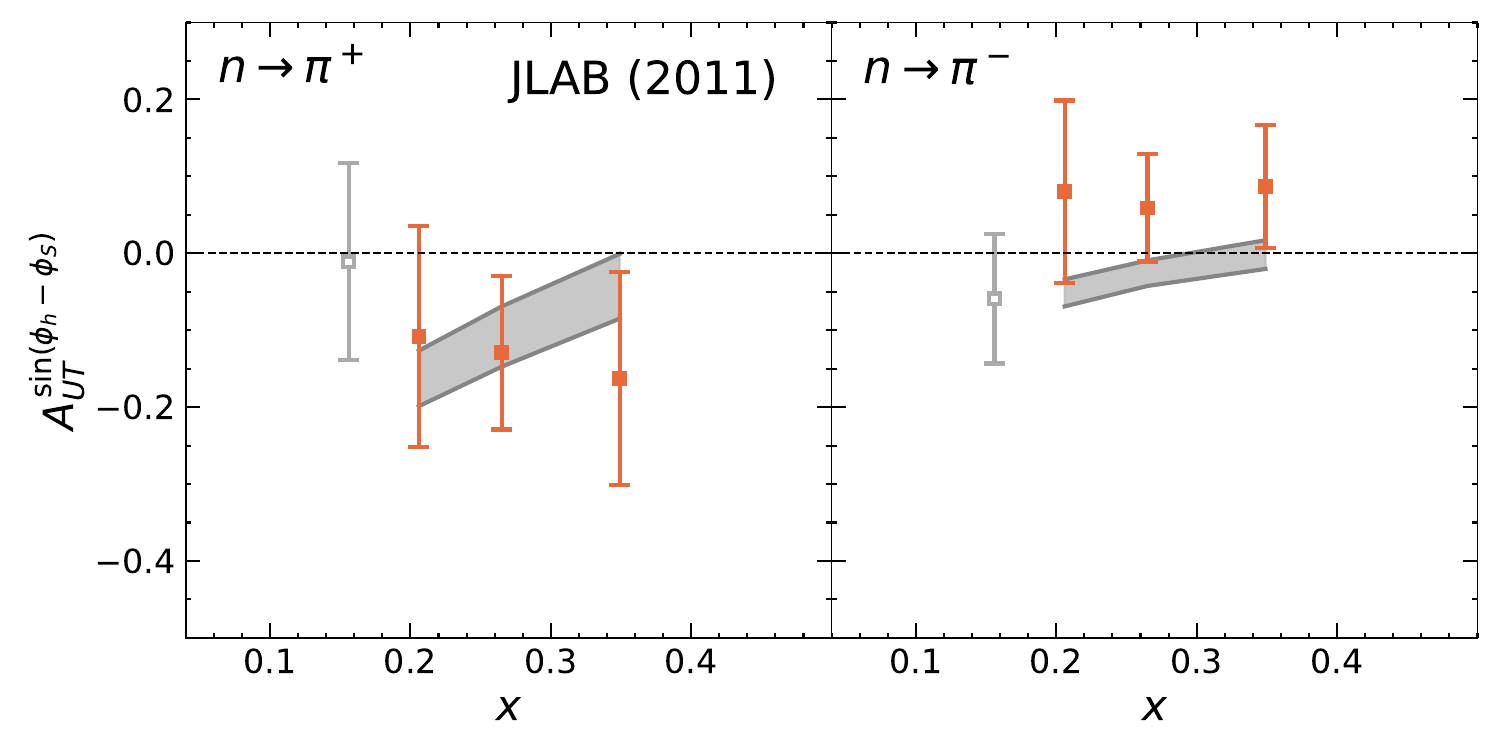}
\end{center}
\caption{JLab Sivers asymmetries from SIDIS off a deuteron target ($^6$LiD)
  with production of positive and negative $\pi$ in the final state~\cite{Qian:2011py}, presented as function of $x$. Same notation as in Fig.~\ref{f:Siv_C09}. 
  %Only the $x$-dependent projections have been included in the fit.
  } 
\label{f:Siv_JLab}
\end{figure}

\begin{figure}[h!]
\begin{center}
\includegraphics[width=\textwidth]{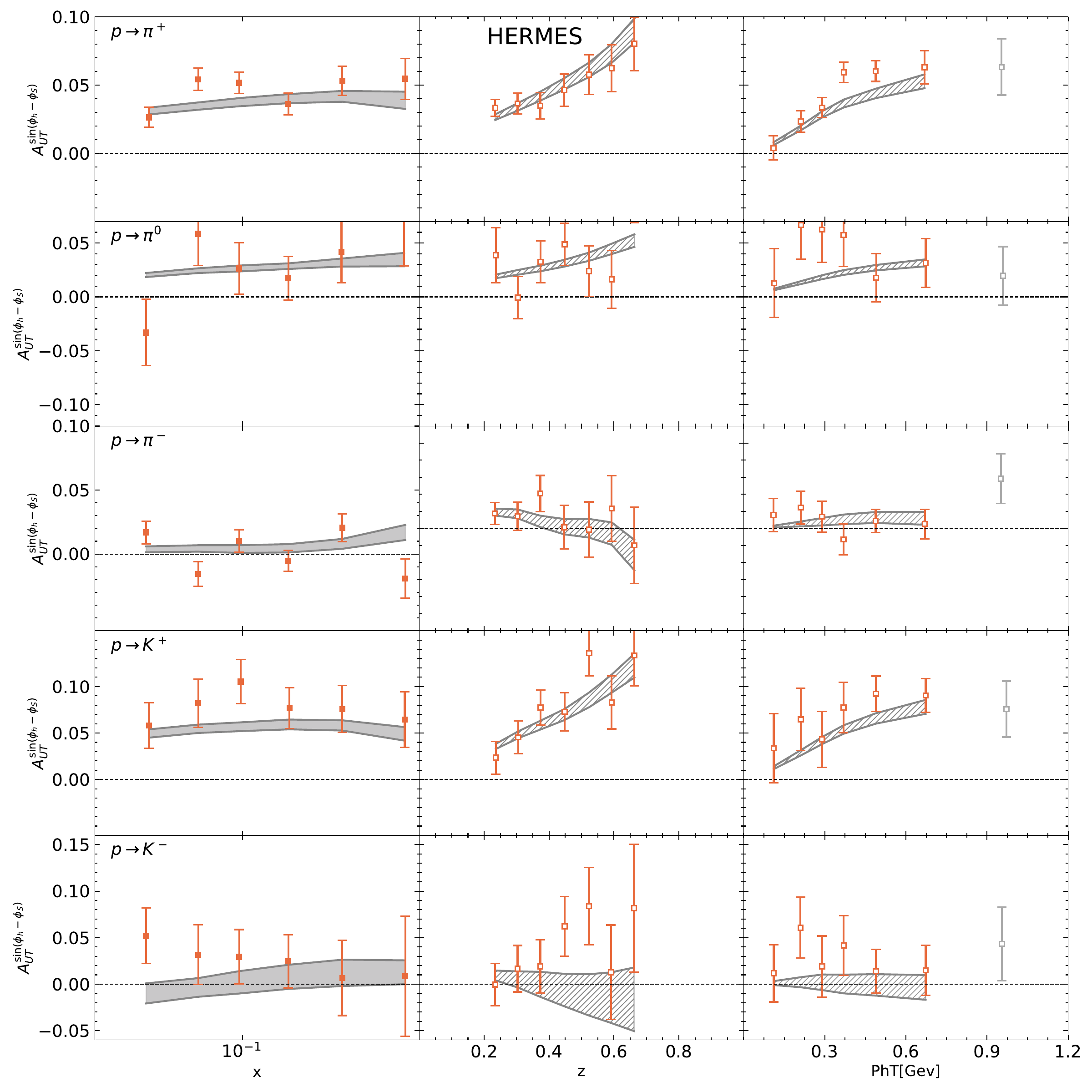}
\end{center}
\caption{HERMES Sivers asymmetries from SIDIS off a proton target (H) with
  production of $\pi^+$, $\pi^0$, $\pi^-$, $K^+$, $K^-$ in the final state~\cite{Airapetian:2009ae}, presented as a function of $x$, $z$, $P_{hT}$. Same notation as in Fig.~\ref{f:Siv_C09}. } 
\label{f:Siv_Hermes}
\end{figure}

\begin{figure}[h!]
\begin{center}
\includegraphics[width=\textwidth]{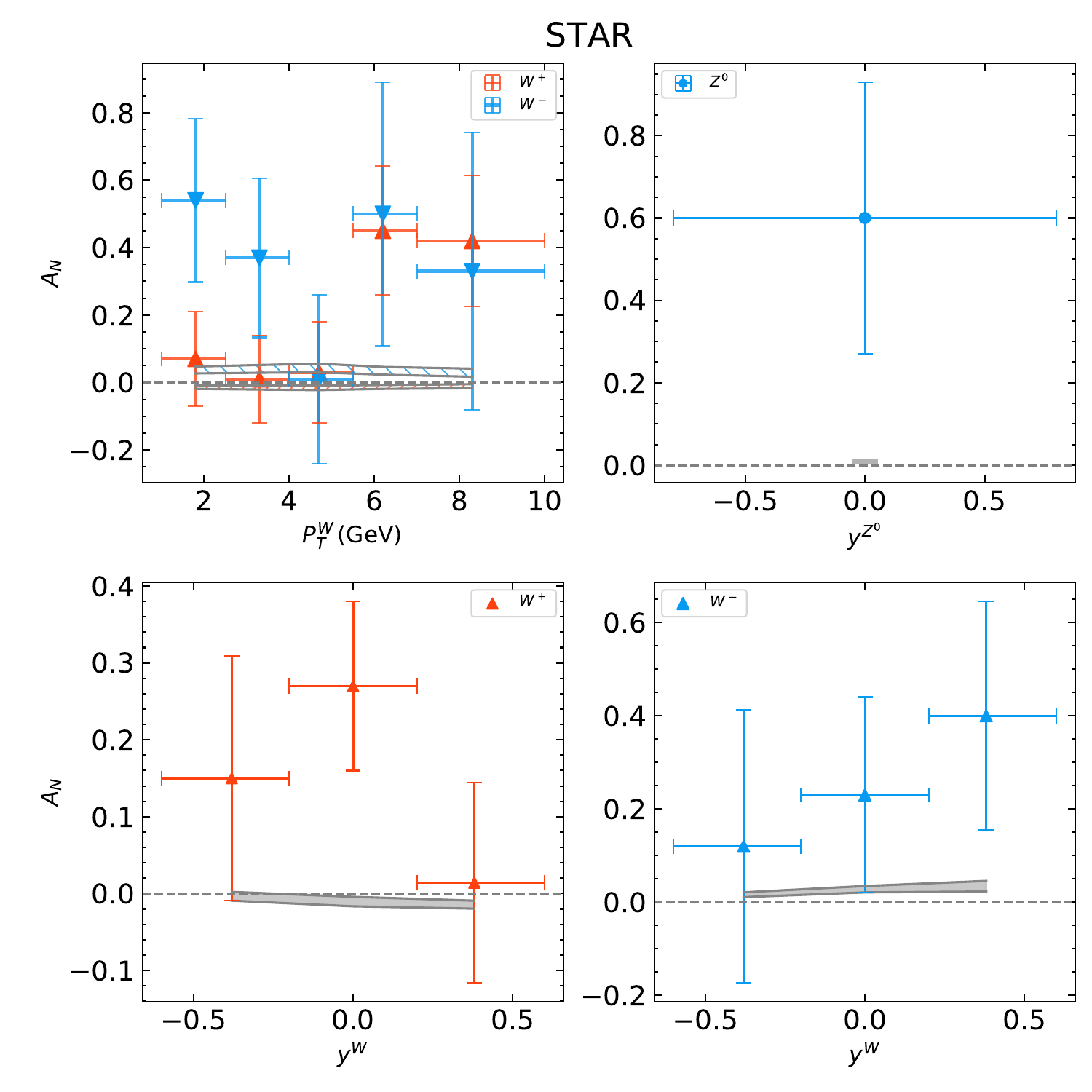}
\end{center}
\caption{STAR Sivers asymmetries from $p - p^\uparrow$ collisions producing $W^\pm$ and $Z^0$ in the final state~\cite{STAR:2015vmv}, presented as function of rapidity $y$ and $p_T^W$. Same notation as in Fig.~\ref{f:Siv_C09}.} 
\label{f:Siv_STAR}
\end{figure}

In Fig.~\ref{f:Siv_C09}, we report the results for the {\tt COMPASS} 2009 run~\cite{Alekseev:2008aa} (32 points with $\chi^2 = 26.5 \pm 4.2$), while in Fig.~\ref{f:Siv_C17piu} and Fig.~\ref{f:Siv_C17meno} we show the 2017 run for positive and negative final state hadrons, respectively~\cite{Adolph:2016dvl} (50 points with $\chi^2 = 31.3 \pm 3.8$). The results for JLab~\cite{Qian:2011py} are depicted in Fig.~\ref{f:Siv_JLab} (6 points with $\chi^2 = 4.1 \pm 0.7$). The {\tt HERMES} results~\cite{Airapetian:2009ae}, together with predictions of projections on different variables, are shown in Fig.~\ref{f:Siv_Hermes} (30 points with $\chi^2 = 46.2 \pm 3.8$). Finally, Fig.~\ref{f:Siv_STAR} contains the results for $W^\pm$ and $Z^0$ production measured by the STAR Collaboration~\cite{STAR:2015vmv} (7 points with  $\chi^2=13.97\pm 0.6$). 

%Finally in Fig.~\ref{f:xf1T1perp} we have the curves for $xf^{(1)}_{T1\perp}(x;Q^2)$ for up, down and sea quarks

%\begin{figure}
%\begin{center}
%  \includegraphics[width=0.5\textwidth]{img/TROTH_xf1Tperp1_all+banda_interp=False.pdf}
%\begin{tabular}{c}
%  \includegraphics[width=0.5\textwidth]{xf1Tperp1_u_all+banda_noylabel_interp=False}
%  \\
%  (a)
%  \\
%%  \includegraphics[width=0.5\textwidth]{xf1Tperp1_d_all+banda_noylabel_interp=False}
%  \\
%  (b)
%  \\
%  \includegraphics[width=0.5\textwidth]{xf1Tperp1_s_all+banda_noylabel_interp=False}
%\\
%%(a) & (b) &
%(c)
%\end{tabular}
%\end{center}
%\caption{The first transverse moment of the Sivers function,
%  $xf_{1T}^{\perp  (1)}$,  as a function of $x$ calculated
%  for the up, down and sea quarks at the scale $Q^2=4$
%  GeV$^2$. The plots show all the 200 replicas obtained from the fit.
%  For each value of $x$, the uncertainty bands contain the central
%  $68\%$ of the replicas.} 
%\label{f:xf1T1perp}
%\end{figure}

\bibliography{SiversFit.bib}
\bibliographystyle{myrevtex}

\end{document}